\documentstyle[12pt,fleqn]{article}
\def\beq {\begin{eqnarray}}
\def\eeq {\end{eqnarray}}
\def\beqn {\begin{eqnarray*}}
\def\eeqn {\end{eqnarray*}}

\def\ni {\noindent}

\def\PL #1 #2 #3 {Phys. Lett.~{\bf#1} (#2) #3}
\def\NP #1 #2 #3 {Nucl. Phys.~{\bf#1} (#2) #3}
\def\ZP #1 #2 #3 {Z.~Phys.~{\bf#1} (#2) #3}
\def\PR #1 #2 #3 {Phys. Rev.~{\bf#1} (#2) #3}
\def\PP #1 #2 #3 {Phys. Rep.~{\bf#1} (#2) #3}
\def\PRL #1 #2 #3 {Phys. Rev.~Lett.~{\bf#1} (#2) #3}

\def\GeV{\mbox{GeV}}

\begin{document}
\begin{titlepage}
%
\vspace*{0.3cm}
\baselineskip=0.7cm
\begin{center}
{\Large \bf Off-shell pion structure function and\\
                flavor asymmetry in the nucleon sea}
\end{center}
%
\vskip 0.9cm
\centerline{\large
        Takayuki Shigetani$^{a)}$\footnote{
                \ni{e-mail address : shige@atlas.phys.metro-u.ac.jp}},
        Katsuhiko Suzuki$^{b)}$\footnote{
                \ni{e-mail address :
                ksuzuki@physik.tu-muenchen.de}}
              and
        Hiroshi Toki$^{c)}$
}
\vskip 0.7cm
\centerline{$^{a)}$\em Department of Physics,
Tokyo Metropolitan University}
\centerline{\em Hachiohji, Tokyo 192-03, Japan}
\vskip 0.3cm
\centerline{$^{b)}$\em Institut f\"{u}r Theoretische Physik,
Technische Universit\"{a}t M\"{u}nchen}
\centerline{\em D-85747 Garching, Germany}
\vskip 0.3cm
\centerline{$^{c)}$\em Research Center for Nuclear Physic (RCNP),
Osaka University}
\centerline{\em Ibaraki, Osaka 567, Japan}
\vskip 1.3cm
\baselineskip=0.7cm
\centerline{\bf ABSTRACT}
\vspace{0.1cm}
  We study off-shell effects of the pion cloud on the sea quark
distribution in the nucleon.
The structure function of the off-mass-shell pion is obtained within
the Nambu and Jona-Lasinio model, where the shape of the distribution
function depends on the pion momentum.
By using the momentum dependent pion structure function, the SU(2) flavor
asymmetry in the nucleon see is reexamined within the pionic model,
in which
the sea quark distribution of the nucleon is given as a convolution of
the off-shell pion structure function.
We calculate experimentally observed quantities related to the SU(2)
asymmetry of the nucleon sea
such as the Gottfried sum and the Drell-Yan cross section
ratios, and find these quantities are insensitive to the off-shell
effects of the pions structure function.
However, the $x$ dependence of
$\bar{u}-\bar{d}$ calculated with the off-shell structure function shows
a clear deviation from the result with the on-shell one.
We evaluate also the cross section of semi-inclusive
lepton-nucleon process with a slow nucleon production to demonstrate
the off-shell effect to be checked experimentally.

\vspace{0.5cm}
\baselineskip = 0.5cm
PACS numbers: 13.60.Hb, 14.40.Aq, 24.85.+p, 12.39.Fe, 12.39.Ki

\end{titlepage}
%
\baselineskip=0.9cm
\ni
{\bf 1. Introduction}

  Recently, many interesting experiments in high energy region, e.g.
deep inelastic lepton-hadron scattering, have been carried out to study
detailed structure of hadrons.
In particular, a progress of experimental technique makes it possible
to investigate the structure function at very small Bjorken $x$,
and thus much theoretical and experimental interests are devoted to the
study of the sea quark distribution of the nucleon.

  The New Muon Collaboration (NMC) \cite{NMC} vigorously observed the
nucleon structure function at the small $x$ in which the sea quark is
dominant.  Assuming that there are equal numbers of $u \bar{u}$ and $d
\bar{d}$ pairs in the nucleon sea, the difference of the neutron proton
structure functions $F_{2}^{ep}(x)-F_{2}^{en}(x)$
gives a value of 1/3, which is so called as the Gottfried sum rule.
However, the experimentally observed value is substantially smaller
than 1/3, which indicates $\bar d$ excess in the nucleon sea, i.e.
the violation of SU(2) flavor symmetry.

In order to confirm the large SU(2) flavor asymmetry of the nucleon sea,
the E772 Collaboration measured the cross section of $p$-$A$
Drell-Yan (DY) process\cite{E772}.
They use both an isoscalar and a neutron excess nuclei as targets,
and extract the information on the
flavor symmetry breaking in the nucleon sea.
Although the data have a large amount of error,
there is no evidence for the SU(2) flavor asymmetry of the nucleon sea.
Very recently, NA51 Collaboration also observed another quantity,
$A_{DY}=\frac{\sigma_{pp}-\sigma_{pn}}{\sigma_{pp}+\sigma_{pn}}$,
by using the $p$-$p$ and $p$-$n$ DY processes \cite{NA51}.
The measured value for $A_{DY}$ indicates the nucleon sea has the strong
isospin asymmetry; $A_{DY}=-0.09\pm 0.025$ at $x=0.18$.
More precise experiments are necessary to get correct understandings of
the nucleon sea structure.

On the other hand, many theoretical works were done
to explain the SU(2) flavor asymmetry of the nucleon
sea\cite{Feynman,Kumano,Henley,others}.
Some authors studied this problem within
``the pionic model"\cite{Kumano,Henley}.
The pionic model is based on the concept that a part of the sea quark of
the nucleon consists of the virtual pion cloud around the nucleon.
One gets the sea quark distribution function of the nucleon
as a convolution of the quark distribution in the pion with the pion
cloud distribution in the nucleon\cite{Sullivan}.
Consequently, $\bar{d}$ excess of the proton sea occurs
according to the processes $p \rightarrow \pi^{+}+n$, $\pi^{0}+p$.
About a half of the NMC experimental
value is reproduced within the pionic model\cite{Kumano}.

In those studies, the pion structure function obtained
by the analysis of the $\pi$-$p$ DY experimental data is employed to
calculate the convolution integral.
However, the virtual pion cloud surrounding the nucleon is essentially
off-mass-shell, and hence its structure is expected to be different from
the structure of the on-shell pion observed in the DY experiment.
If so, it is important to examine how the off-shell pion structure
function differs from the on-shell one, and
how such a difference
affects some observations in high energy experiments.

In the recent work of the present authors\cite{Shige2}, the off-shell pion
structure function was calculated in the Nambu and Jona-Lasinio (NJL)
model, where the off-shell pion structure function shows a substantial
momentum dependence.
In this paper, we shall study the sea quark structure of the nucleon and
its
isospin asymmetry in terms of the off-mass-shell pion structure function.
At first, we briefly show the procedure to calculate the off-shell pion
structure function in the NJL model.
Using the momentum dependent structure function, we then evaluate various
quantities associated with the nucleon sea quark distribution within
the pionic
model.   Effects of the off-shell pion on the SU(2) flavor asymmetry of the
nucleon sea are discussed.
We also examine the semi-inclusive deep inelastic
lepton-nucleon scattering with a slow nucleon in the final
state, which was proposed in Ref.~\cite{Shige2,Semi1,Semi2,Semi3},
to observe the off-shell pion structure function directly.
All the results are compared with those calculated by using the on-shell
pion structure function obtained experimentally.

This paper is arranged as follows.  We show the derivation of the
off-mass-shell pion structure function in Section 2.  In Section 3,
we briefly introduce the pionic model, and calculate
$F^p_2(x) - F^n_2(x)$, from which
we can extract the violation of the isospin symmetry in the nucleon sea.
We also calculate the DY cross section ratios in Section 4, which
give alternative information of the nucleon sea.
Section 5 is devoted to the study of the semi-inclusive process to
clarify the off-shell effect for the pion structure function.
Discussions and conclusions are presented in the final section.

\vspace{4cm}
\ni
{\bf 2. Off-shell pion structure function in the NJL model}

The NJL model is believed to be an effective theory of QCD
at low energies\cite{NJL,NJL2}.
This model possesses chiral symmetry, which is one of the most important
aspects of the low energy QCD, and demonstrates the spontaneous
breakdown of the chiral symmetry.
By solving the Dyson equation for the quark propagator
in the Hartree approximation, we get the constituent quarks
with dynamical masses.
One can also obtain the meson modes by
solving the Bethe-Salpeter (BS) equation with the constituent quarks.
Once the model parameters are fixed to reproduce the pion mass and
the decay constant, the NJL model describes SU(3) meson properties
well\cite{NJL2}.

Recently, several attempts were done to calculate the hadron structure
function in term of the low energy effective theory of
QCD\cite{JaffeRoss}.
The basic procedure is summarized as follows\cite{JaffeRoss};
We first evaluate the twist-2 matrix element, which is a leading term of the
light-cone expansion of the forward scattering amplitude, within the QCD
effective theory.
Higher twist terms are safely neglected, since we are only interested in the
Bjorken limit.
By doing so, we get the structure function at the low energy
scale $Q^2_0$.
Here, the low energy model scale is assumed to be about
1 GeV or less, at which the effective theories are supposed to work.
After the calculated structure function is evoluted to the experimental
momentum scale in terms of the Altarelli-Parisi equation\cite{AP},
we can compare it with the experimental data at high energies.
The structure function obtained by the effective theory provides a boundary
condition at the low energy scale $Q^2_0$ for the QCD evolution equation.

In the previous studies of the present authors, the structure functions of
mesons is evaluated in the NJL model\cite{Shige1,Shige2}.
We compute the forward scattering amplitude in the impulse approximation.
We only deal with a leading term of $1/Q^2$ expansion, which coincides with
the twist-2 contribution and is enough for our purpose\cite{Rujula}.
Since the the hadronic tensor is related with the forward scattering amplitude
through the optical theorem\cite{Muta},
we can obtain the quark distribution function at the model
scale.

Here, we are interested in the off-mass-shell behavior of the
pion structure function.
We can get the off-shell pion structure function within the
NJL model in the same way as for the on-shell pion, though we do not know
whether or not the NJL model works for the off-shell pion properties.
We find a following expression for the quark distribution function of the
off-shell pion at the low energy scale\cite{Shige2},
%
\begin{eqnarray}
V_{\pi}(x,t) & \propto & -g_{\pi qq}^{2} N \int_{-\infty }^0 \! d\mu^{2}
 [\frac{1}{\mu^{2}-m^{2}}+x\frac{t}{(\mu^{2}-m^{2})^{2}}]\nonumber\\
\jot 2cm
& & \hspace {3cm} \times \theta (-t x(1-x)-xm^{2}-(1-x)\mu^{2}) \; .
\label{pi(x)}
\end{eqnarray}
\noindent
where $x$ is the Bjorken-$x$, which means the longitudinal momentum fraction
carried by the quark.
Here, \(\theta\) is the usual step function, and $g_{\pi qq}$
the pion-quark-quark coupling obtained by the NJL model.
$\mu^{2}$ is the invariant mass of the struck quark and
$m$ the constituent quark mass.
Note that the expression (\ref{pi(x)}) of the structure function for
the off-shell case differs from that of the on-shell case only in the
use of $t$ instead of $-m_{\pi}^{2}$.
$t (= - p_{\pi}^{2})$ indicates the pion
4-momentum squared, which is $p_{\pi}^{2}=m_{\pi}^{2}$ for the on-shell pion
case.
The NJL model is a non-renormalizable theory, and thus
requires the finite momentum cutoff \(\Lambda\sim 1\) GeV,
which is identified with the typical scale of
the chiral symmetry breaking.
We introduce the Fermi-distribution type momentum cutoff
function in the equation (\ref{pi(x)}) \cite{Shige2}.
The choice of the cutoff function dose not alter
the behavior of the off-shell pion structure function (\ref{pi(x)}) largely.

Equation (\ref{pi(x)}) provides the valence quark distribution
function of the off-shell pion at the low energy model scale,
which is taken to be $Q_{0} = 0.5{\GeV}$.
After carrying out the $Q^2$ evolution with the help of the Altarelli-Parisi
equation,
we show in Fig.1 the momentum, $t(=-p^2_\pi > 0)$, dependence of the pion
structure function at $Q^2 = 20{\GeV}^2$ with the on-shell
one ($p_{\pi}^{2}=  m^2_\pi$).
As the pion momentum $t$ increases,
the peak position of the quark distribution moves toward the small $x$
region, and the distribution function shows a substantial reduction
for $x>0.4$.
Around $x \sim 0.5$, the absolute value of the
distribution function at $t = 0.5{\GeV}^2$ is almost half of that for the
on-shell case.

\vspace{4cm}
\ni
{\bf 3. Pionic model and isospin asymmetry of the nucleon sea}

The pionic contribution is one of the candidates to explain
the iso-spin asymmetry of nucleon sea \cite{Kumano,Henley,others}.
A proton has the pion cloud due to the process
$p \rightarrow n+\pi ^{+}$.
Since this process is equal to ``$u \rightarrow d(u \bar{d})$''
in the sense of the quark model, it produces an excess of
$\bar{d}$ over $\bar{u}$ in the proton sea.
On the contrary, the process $p \rightarrow \Delta ^{++} +\pi ^{-}$
produces the excess of $\bar{u}$ over $\bar{d}$, which contributes to
cancel the $\bar{d}$ excess.
It is very important to take into consideration both processes, which
are shown schematically in Fig.2(a) and (b).

The pionic contribution to the sea quark distribution function in the nucleon
is given by the convolution integral of the pion momentum
distribution $f_{\pi}^{'}(y,t)$ with
the off-shell pion structure function $\bar{q}_{\pi}(x,Q^{2},t)$:
\begin{eqnarray}
x \bar{q}_{N}^{\pi NN}(x,Q^{2})
=\int^{1}_{x}dy \int^{\infty}_{t_{min}} dt\;
[f_{\pi}^{'(\pi NN)}(y,t) +f_{\pi}^{'(\pi N \Delta)}(y,t)]
\left(\frac{x}{y}\right) \bar{q}_{\pi}(x/y,Q^{2},t) \; ,
\label{convol}
\end{eqnarray}
\noindent
where we adopt the notation used in Ref.~\cite{Kumano}.
Using the one pion exchange approximation,
the pion cloud distribution for Fig.2(a) and (b) is obtained as following:
\begin{eqnarray}
f_{\pi}^{'(\pi NN)}(y,t)&=&I_{\pi NN} \frac{g_{\pi NN}^{2}}{16\pi^{2}}
           \frac{t}{(t+m_{\pi}^{2})^{2}}[F_{\pi NN}(t)]^{2} \\ \nonumber
f_{\pi}^{'(\pi N\Delta)}(y,t)&=&I_{\pi N\Delta}
\frac{g_{\pi N\Delta}^{2}}{16\pi^{2}}
           \frac{(m_{N}+m_{\Delta})^{2}+t}{(t+m_{\pi}^{2})^{2}}
           (\frac{(m_{N}^{2}-m_{\Delta}^{2}+t)^{2}}{4m_{\Delta}^{2}}+t)
           [F_{\pi N\Delta}(t)]^{2} \; .
\label{pimom_dist}
\end{eqnarray}
\noindent
where $I_{\pi NN}$ and $I_{\pi N\Delta}$ are the isospin factors given
in Ref.~\cite{Kumano}.
We also used the dipole cutoff parameter in Ref.~\cite{Kumano},
$\Lambda =0.8 \GeV$, for both the $\pi NN$ and the $\pi N\Delta$ form
factors, $F_{\pi NN}(t),F_{\pi N\Delta}(t)$.
Note that the quark distribution function explicitly depends on the pion
momentum $t$ in equation (\ref{convol}).
As mentioned in Sec. 2, the off-shell pion structure function
$\bar{q}_{\pi}(x,Q^{2},t)$
is obtained in the framework of the NJL model as shown in Fig.1.

We try to apply the pionic model (\ref{convol}) to the DIS and DY
processes.
At first, we pay attention to the NMC experimental data \cite{NMC},
which show the SU(2) flavor symmetry breaking of the nucleon
sea.  Let us define $I_G$ as a integral of the difference
between the neutron and proton structure functions;
\begin{eqnarray}
I_{G}\equiv \int_{0}^{1}\frac{dx}{x}\;[F_{2}^{ep}(x)-F_{2}^{en}(x)] \; .
\label{gsr}
\end{eqnarray}
\noindent
The nucleon structure function is written by the
up-quark distribution function $u(x)$, down-quark $d(x)$ and the
strange-quark $s(x)$, and the corresponding as antiquark distributions,
\begin{eqnarray}
F_{2}^{ep}(x)&=&\frac{4}{9}x[u(x)+\bar{u}(x)]
             +\frac{1}{9}x[d(x)+\bar{d}(x)+s(x)+\bar{s}(x)] \nonumber \\
F_{2}^{en}(x)&=&\frac{4}{9}x[d(x)+\bar{d}(x)]
             +\frac{1}{9}x[u(x)+\bar{u}(x)+s(x)+\bar{s}(x)]
\label{parton}
\end{eqnarray}
\noindent
Here, we assume the iso-spin symmetry of the quark distribution between
a proton and a neutron as usual.
Then, $I_G$ in (\ref{gsr}) can be rewritten as,
\begin{eqnarray}
I_{G}&=&
\frac{1}{3}\int_{0}^{1}\;[u(x)-d(x)]dx
     +\frac{1}{3}\int_{0}^{1}\;[\bar{u}(x)-\bar{d}(x)]dx \nonumber \\
&=&
\frac{1}{3}\int_{0}^{1}\;[u_{v}(x)-d_{v}(x)]dx
     +\frac{2}{3}\int_{0}^{1}\;[\bar{u}(x)-\bar{d}(x)]dx \nonumber \\
     &\equiv&\frac{1}{3} + \Delta I_{G} \; .
\label{gsr2}
\end{eqnarray}
\noindent
where $u_{v}(x)$ and $d_{v}(x)$ are the valence $u$- and $d$-quark
distribution functions, respectively.
If the nucleon sea has the SU(2) flavor symmetry, $\bar u(x) = \bar d(x)$,
$I_{G}$ is equal to 1/3, which is just the Gottfried sum rule.
The NMC data, however, give $I_{G}=0.235\pm 0.026$\cite{NMC}, which
indicate the sea quark violates the SU(2) flavor symmetry.

We shall calculate the difference of the sea quark distribution
($ \bar{u}(x)-\bar{d}(x)$)
of the nucleon with the pionic model.
{}From eq.(\ref{convol}), $\bar{u}(x)-\bar{d}(x)$ is written as,
\begin{eqnarray*}
x [\bar{u}(x)-\bar{d}(x)]^{\pi N}_{N}&=&
\int^{1}_{x}dy \int^{\infty}_{t_{min}} dt\; \\ \nonumber
& &\times [f_{\pi}^{'(\pi NN)}(y,t) +f_{\pi}^{'(\pi N \Delta)}(y,t)]
\left(\frac{x}{y}\right) [\bar{u}(x)-\bar{d}(x)]_{\pi}\; .
\label{fasym}
\end{eqnarray*}
\noindent
The antiquark distribution function of pion is given by both the
valence and sea quark distribution in the pion, ($V_{\pi}(x),S_{\pi}(x)$)
as following\cite{Kumano},
\begin{eqnarray*}
\bar{u}_{\pi^{-}}(x)&=&\bar{d}_{\pi^{+}}(x)\equiv
V_{\pi}(x)+S_{\pi}(x)\nonumber\\
\bar{u}_{\pi^{+}}(x)&=&\bar{d}_{\pi^{-}}(x)\equiv S_{\pi}(x)\nonumber\\
\bar{u}_{\pi^{0}}(x)&=&\bar{d}_{\pi^{0}}(x)\equiv V_{\pi}(x)/2+S_{\pi}(x)
\end{eqnarray*}
\noindent
Thus, we obtain $(\bar{u}-\bar{d})_{\pi^{+}}=-V_{\pi}$ ,
$(\bar{u}-\bar{d})_{\pi^{0}}=0$, and $(\bar{u}-\bar{d})_{\pi^{-}}=V_{\pi}$.

Calculating (\ref{convol}) and summing up $\pi^+$, $\pi^-$ and $\pi^0$
contributions,
we show in Fig.3  the $x$ dependence $[\bar{u}-\bar{d}]_{N}$ by the solid
curve.   Here, we use the off-shell momentum dependent structure function
for $V_\pi(x)$.
For comparison, we also show in Fig.3 the previous result of the pionic
model by the dashed curve, where the on-shell pion structure function is
used for $V_\pi(x)$.  The dash-dotted curve denotes the NMC experimental
fit\cite{PRS}, which gives  a large value for  $[\bar{u}-\bar{d}]_{N}$.
The absolute value of the NMC fit at the peak position is larger
than others of the pionic model, which is already found in the previous
works\cite{Kumano,others}.

For $x<0.1$, the resulting shape of sea quark distribution shows a
clear difference between off-shell(solid) and on-shell (dashed) results, i.e.
off-shell results is large.
In turn, for $x > 0.3$, the result with the off-shell pion
structure function
is much smaller than that with the on-shell one.
This behavior is understood as follows.
The difference $\bar{u}-\bar{d} (x) $,
is calculated by the convolution of the pion
structure function as represented in equation (\ref{convol}).
Recall that the off-shell distribution function for large-$x$ region
is smaller than the on-shell one, and becomes larger for small $x$ region,
as shown in Fig.1.
Hence, as a result of the convolution integral,
$\bar{u}(x)-\bar{d}(x)$ calculated with the off-shell
distribution function becomes smaller at $x>0.3$ region, as compared with
the result with the on-shell distribution.

To estimate the value of $\Delta I_{G}$ which indicates the
violation of GSR, we integrate the
$\bar{u}-\bar{d}$ over $x$ from 0 to 1.
\begin{eqnarray*}
\Delta I_{G} &=& -0.0557 \hspace {1cm} \mbox{: for the on-shell
pion}\nonumber\\
\Delta I_{G} &=& -0.0586 \hspace {1cm} \mbox{: for the off-shell pion}
\end{eqnarray*}
\noindent
Our result with the off-shell pion structure function gives
almost the same value as the one with the on-shell result, though the shapes
between the on-shell and the off-shell cases are somewhat different.
We remark that the value of our
results explains about 50 \% of the discrepancy from the NMC data.
This result is consistent with the previous studies\cite{Kumano}.
The calculated results are insensitive to the cutoff parameter
in the dipole form factors, when it is around $\Lambda \sim 1$ GeV.
The $\pi NN$ and the $\pi N\Delta$ form factors depend on the dipole cutoff
parameter.  The total value of $\Delta I_{G}$, however, does not changed so
much
by the cutoff parameter, because the effects of the cutoff parameter
in the nucleon process and the delta process cancel each other.


\vspace{4cm}
\noindent
{\bf 4 Flavor asymmetry in the Drell-Yan process}

The Drell-Yan process is a
useful tool to separate the valence quark and the sea quark distribution
of the nucleon.
In this process, a quark and an antiquark annihilate to produce a lepton
pair via a virtual photon.
By choosing the suitable kinematical condition,
one can get directly the information of the sea quark
distribution through the DY process\cite{DrellYan2}.

The E772 Collaboration\cite{E772} compares the DY data from
iso-scalar targets ($^{2}\mbox{H}$ and $\mbox{C}$) with data from the
neutron excess nucleus  ($\mbox{W}$).
The ratio of the DY cross sections can be expressed as,
\begin{eqnarray}
R_{W}(x)\equiv \frac{\sigma _{W}(x)}{\sigma _{IS}(x)}
&=&\frac{N_{W}\sigma _{n} + Z_{W}\sigma _{p}}{N_{IS}\sigma _{n} +
Z_{IS}\sigma _{p}}  \nonumber \\
\jot 1cm
&\simeq&1+\frac{N-Z}{A}\cdot \frac{\bar{u}(x)-\bar{d}(x)}
{\bar{u}(x)+\bar{d}(x)}\nonumber\\
\jot 1cm
&=& 1+0.183\cdot \frac{\bar{u}(x)-\bar{d}(x)}{\bar{u}(x)+\bar{d}(x)} \;\;\; ,
\label{ratio_w}
\end{eqnarray}
\noindent
where $\sigma$ is the DY cross section per nucleon, $\mbox{IS}$ stands
for iso-scalar, and N, Z refer to a target with a neutron excess.
If this ratio deviates from 1, the SU(2) flavor asymmetry exists in the
nucleon sea.

The experimental data of E772 is shown in Fig.4.
We can not find the large SU(2) flavor asymmetry of the nucleon in this
experimental data, though the data have a large amount of error.
We calculate the $R_{W}$ using the pionic model.
One also requires knowledge of the sea quark distribution function,
$S_{\pi}(x)$, of the pion to get the shape of
$\bar{u}+\bar{d}$ (see Section3).
We simply use the parameterization of Sutton {\it et al.}\cite{SMRS},
because we do not evaluate the sea quark distribution of
the pion in the NJL model,

We show in Fig.4 the calculated results obtained in the pionic
model with on-shell and off-shell pion structure functions by the
solid and the dashed curve, respectively.
We also show the calculated results using the NMC experimental fit for $\bar u
(x)$ and $\bar d (x)$, which is depicted by the dash-dotted curve.
The magnitude of the pionic model results are smaller than the result
calculated by the NMC fit.
The SU(2) flavor asymmetry of the nucleon in the pionic model
is weak compared with the result described by dot-dashed curve, as we
already discussed in section 3.
The E772 experimental data, however, have a large amount of error.
Both our results in the pionic model and the one calculated by the NMC fit
are consistent to the experimental data
within the range of the experimental error.

On the other hand, we can not find a clear evidence of the off-shell effect
in Fig.4.  The pionic model results with the on-shell and off-shell pion
structure functions are almost the same.
The situation is bit different from the result in Fig.3, where
the $[\bar{u}-\bar{d}](x)$ has relatively distinct
off-shell effect.
In this case, we take a ratio of the DY cross section.
Here, there are off-shell effects in both the
denominator and numerator in equation (\ref{ratio_w}).
Namely, the off-shell effect is partially canceled
by taking a ratio in equation (\ref{ratio_w}), and thus
not evident in the calculated results.

There is also a recent DY result by the NA51 Collaboration \cite{NA51}.
They measure the $p$-$p$ and $p$-$n$ DY processes, and extract
$A_{DY}$ to study the flavor asymmetry of the nucleon sea.
\begin{eqnarray}
A_{DY}(x)&\equiv& \frac{\sigma _{pp}(x)-\sigma _{pn}(x)}
                     {\sigma _{pp}(x)+\sigma _{pn}(x)} \nonumber\\
&\simeq& \frac{(4u_{v}-d_{v})(\bar{u}-\bar{d})+(4\bar{u}-\bar{d})(u_{v}-d_{v})}
{(4u_{v}+d_{v})(\bar{u}+\bar{d})+(4\bar{u}+\bar{d})(u_{v}+d_{v})}
\label{ratio_dy}
\end{eqnarray}
\noindent
We show in Fig.5 three calculated results using the sea quark distribution
from NMC fit (dot-dashed curve) and pionic model with the on-shell
(dashed curve) and the off-shell (solid curve) pion structure function.
The result with the off-shell structure function is not so different
from the one with the on-shell structure function.
This is because the off-shell effect is canceled by taking a ratio, as
mentioned before.
We see the pionic model results with both the off-shell and the
on-shell pion structure
functions do not fully account for the experimental data, which indicate the
large SU(2) flavor asymmetry.

\vspace{4cm}
\ni
{\bf 5. Semi-inclusive process with a slow nucleon production}

We calculate here the cross section of the semi-inclusive process
with a slow nucleon production\cite{Semi1,Semi2,Semi3},
using the off-shell pion structure
function in the NJL model.
If we observe a slow nucleon ($p < 1GeV/c$) in the final state,
this semi-inclusive process
is assumed to be dominated by the virtual photon-virtual pion deep
inelastic
scattering.
This cross section has not been observed yet,
but we expect that it contains the information of the off-shell
pion structure function through one pion exchange mechanism.

We focus on the following process;
\begin{eqnarray}
e+p\to e'+ n +X \;.
\label{process}
\end{eqnarray}
\noindent
The cross section
of the electron-proton scattering process with the one
pion exchange is written as\cite{Shige2,Semi2},
\begin{eqnarray}
\frac{d\sigma}{dE'_{L}d \cos \theta_{L}d^{3}p'} &=&
     \frac{4\alpha^{2}}{\pi Q^{4}} \frac{g^{2}_{\pi NN}}{4\pi}
     \frac{E'^{2}_{L}}{m_{N}E} \frac{t}{(t+m_{\pi}^{2}) ^ 2}\nonumber\\
        & & \times \{
 2 \sin^{2}\frac{1}{2}\theta_{L}\,F_{1}^{\pi}(x_{\pi},Q^2 , t)
\nonumber\\
 & &+[\frac{(k\cdot P_{\pi})(k'\cdot P_{\pi})}{E_{L}E'_{L}}
 + t \sin^{2}\frac{1}{2}\theta_{L}]
        \frac{F_{2}^{\pi}(x_{\pi},Q^2 , t)}{q \cdot P_{\pi}}
        \} \; ,
\label{cross}
\end{eqnarray}
\noindent
where $E_{L}$ and $E'_{L}$ are the energy of the initial and the final
electron, respectively.
$\theta_{L}$ is the angle between the initial and the final electron beam
directions.
$p'$ is the 3-momentum of the slow nucleon in the final state.
$P_{\pi}$ is the pion 4-momentum, $m_{\pi}$ the pion mass, and $m_{N}$
the nucleon mass.
Further,
\begin{eqnarray}
        F_{2}^{\pi}(x_{\pi},Q^2 , t) &=&
        2x_{\pi} F_{1}^{\pi}(x_{\pi},Q^2 , t) \nonumber \\
&=&
x_{\pi}\sum_{i}e_{i}^{2}[q_{i}(x_{\pi},t)+\bar{q}_{i}(x_{\pi},t)] \nonumber \\
 &=&  x_{\pi}\sum_{i}e_{i}^{2} [V^i_\pi (x_\pi,t) + 2 S^i_\pi (x_\pi) ]
\label{offpi}
\end{eqnarray}
\noindent
is the off-shell pion structure functions. We apply the Reggeized pion
to $g_{\pi NN}$ vertex \cite{Semi2}.
We can utilize also the dipole form factor for the $g_{\pi NN}$ vertex.
Since the latter case provides almost identical results, we discuss only
the case of the Reggeized pion.

Using (\ref{cross}),
we calculate the cross section with the incident
electron energy $E_{L}=30\GeV$ which is accessible at DESY.
Here, we use the distribution functions of the off-shell pion
at $Q^2 = 20 \GeV^2$ shown in Fig.1.
Detailed methods of the calculation are found in Ref.~\cite{Shige2}.

We calculate the double differential cross section using both the off-shell
and the on-shell pion structure functions.
The absolute value of the double differential cross section is of the
order of $10^{-5}$mb.
Then, we take a ratio $\frac{[d^{2}\sigma /dxdp]_{\mbox{\small{off-shell}}}}
{[d^{2}\sigma /dxdp]_{\mbox{\small{on-shell}}}}$ in order to clarify the
off-shell effect on the pion structure function.
The result is shown in Fig.6(a) as a function of $x$ with the momentum of the
final state $p' $ being fixed for the case of the Reggeized pion.
The resulting ratio shows that the cross section calculated by the off-shell
structure function is much smaller than that of the on-shell pion
for  $x>0.1$.
Note that the contribution from the sea quark $S_\pi(x)$ is dominant
only in small $x$ region.
Hence, this behavior of the ratio is caused by the change of
the valence quark distribution function in the off-shell and
on-shell pion.
Remember that the off-shell pion structure function becomes smaller than
the on-shell one as the momentum $t$ increases, which is already shown in
Fig.1.
Hence, after the convolution integral (\ref{cross}),
the resulting cross section calculated by the off-shell pion shows
substantial reduction for the large $x$ as compared with the on-shell pion
result.

\vspace{4cm}
\ni
{\bf 6. Conclusion}

In this paper, we have studied off-shell effects of the pion cloud
on the sea quark distribution in the nucleon.
One can measure the pion structure function using the $\pi$-$N$ Drell-Yan
process, but the off-shell behavior of the pion
structure function is not yet understood, which might be important to
study the sea quark distribution function of the nucleon.
At first, we have calculated the structure function for the off-mass-shell
pion within the NJL model.
The result clearly depends on the pion momentum square $t$.
As the off-shell pion momentum $t = -p_\pi^2$ increases,
the peak position of the structure function shifts to the small $x$
region, and the distribution function shows a substantial reduction
for $x>0.4$.

Next, we have calculated the sea quark distribution of the nucleon
in the framework of the pionic model.
We suppose that the sea quark in the nucleon is composed of
pion cloud around the nucleon.
The pions around the nucleon are virtual particles, and
thus we have to use the off-shell pion structure function.

We have calculated the SU(2) flavor asymmetry of the nucleon sea quark
distribution, measured in DIS scattering and DY experiments,
using the pionic model with the off-shell pion structure function.
The off-shell behavior of the
pion structure function changes the shape of $\bar{u}(x)-\bar{d}(x)$.
The result with the off-shell pion structure function is much larger than
that with the on-shell one for small $x$, and smaller in the
large Bjorken-$x$ region ($x > 0.3$).
Because of the reduction of the off-shell pion structure function at
the large $x$,
the peak of $\bar{u}-\bar{d}$ with the off-shell pion structure function
shifts toward the small $x$ region as compared with the on-shell one.

To get the value of the violation of GSR, we have integrated out
$\bar{u}-\bar{d}$ by $x$.
Our computed results with the off-shell structure function is almost the
same as the on-shell result of the previous study, though the shape of
$\bar{u}-\bar{d}$ is different.
Off-mass-shell effect of the pion cloud is not apparent in the value of
the GSR, which is an integrated quantity.

We have also calculated the quantities related to the nucleon sea,
which are obtained in the experiments of DY process.
One is the experimental data of E772 and another result is done by the
NA51.
In both cases, we do not find clear difference between the
results with off-shell and the on-shell structure functions.
We can not find the evidence of the off-shell effect in the DY data.
This is because the off-shell effect is partially canceled
by taking a ratio $R_{W}$ and $A_{AD}$.

If we compared our calculations with the DY cross section directly (not the
ratio), we could extract the off-shell effects of the pion.
However,
the DY cross section involves ambiguity about the absolute normalization
factor, $K$-factor.  The $K$-factor is not precisely determined at the moment.
Hence, we avoid direct comparison of our calculation with the DY cross
section.

Finally, we have calculated the cross section of the semi-inclusive
scattering with a slow nucleon production to see the off-shell pion effect.
The calculated results show a clear difference between
the off-shell pion and the on-shell pion.
No such experiments are available at present,
but we expect that it contains the information of the off-shell
pion structure function through the one pion exchange mechanism.
One is able to get the information of the off-shell pion, if
the accurate experiment of this type is carried out.


%
%
\newpage
\noindent
{\bf {\Large Figure Captions\/}}
\vspace {1cm}

\ni
Fig. 1  : The off-shell pion structure function with various pion
momenta at $Q^{2}=20 \GeV^{2}$ as a function of the Bjorken $x$.
We show three cases for off-shell pion momenta
squared $t=-p_{\pi}^{2}$, where the dashed curve is $t= 0.1 \GeV^{2}$,
dotted curve  $t=0.25 \GeV^{2}$,
and dot-dashed curve  $t=0.5 \GeV^{2}$.
We indicate the on-shell pion ($p_{\pi}^{2}=m_{\pi}^{2}$)
structure function for comparison by the solid curve.

\vspace {1cm}

\ni
Fig. 2  : The diagrams for the pionic contributions to the deep
inelastic lepton-nucleon scattering.
Diagrams (a) and (b) express the $\pi NN$ and
$\pi N\Delta$ vertices, respectively.
The solid line denotes the electrons and the thick line the nucleon.
The dashed line indicates the pion and the wavy line the virtual
photon.  The double line is for the nucleon in (a) and for the delta
in (b).

\vspace {1cm}

\ni
Fig. 3  : The difference $x[\bar{u}(x)-\bar{d}(x)]$ as
a function of the Bjorken $x$.
Dot-dashed curve is the one extracted from the NMC experimental fit,
Ref.~\cite{PRS}.
The dashed curve is calculated with the on-shell pion structure function,
and the solid curve with the off-shell one.

\vspace {1cm}

\ni
Fig. 4  : The ratio $R_{W}$ of the DY cross section as a function of
the Bjorken $x$.
Experimental data are taken from E772 Coll \cite{E772}.
Notations for theoretical calculations are the same as those in Fig.3.
See text for details.
\vspace {1cm}

\ni
Fig. 5  : The ratio $A_{DY}$ for DY
production as a function of the Bjorken $x$.
Experimental data are taken from NA51 Coll \cite{NA51}.
Notations for theoretical calculations are the same as those in Fig.3.
See text for details.

\vspace {1cm}

\ni
Fig. 6  : The ratio of cross sections of a lepton nucleon
scattering process with a slow nucleon in coincidence in the final state
with off-shell and on-shell pion structure functions as a function of
Bjorken $x$.
The cross section is calculated by using eq.(\ref{cross}) with
Reggeized pion for the $g_{\pi NN}$ vertex.
The result with the produced proton momentum $p'=1.3GeV/c$ is depicted
by the solid curve, and the one with $p'=0.5 \GeV /c$ by the dotted curve.

\vspace {1cm}

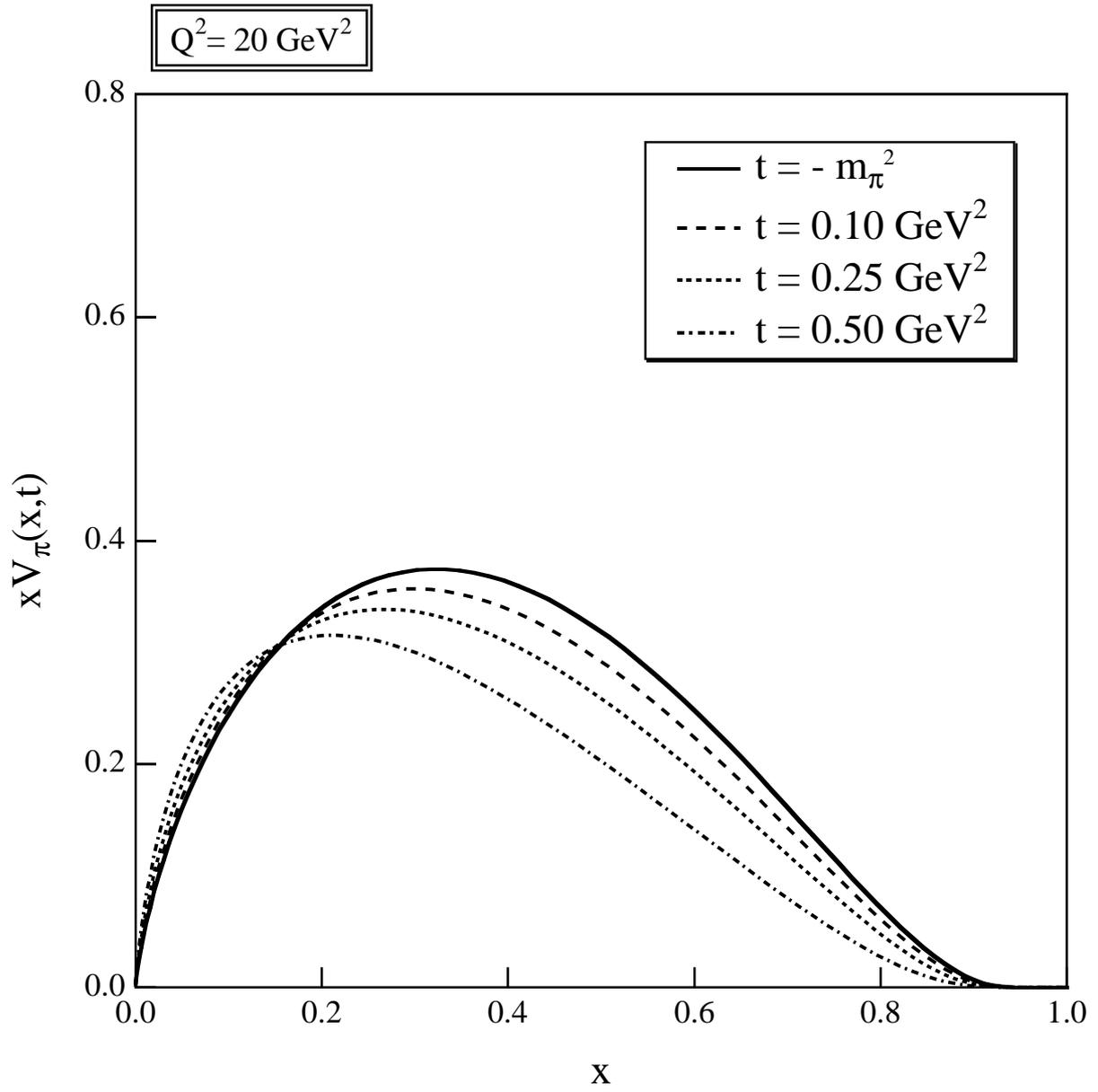
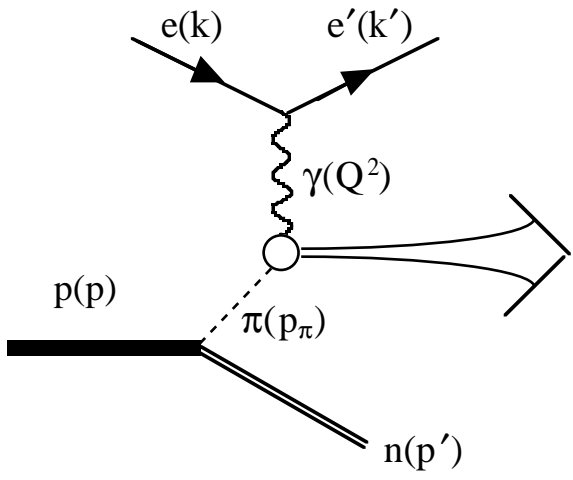
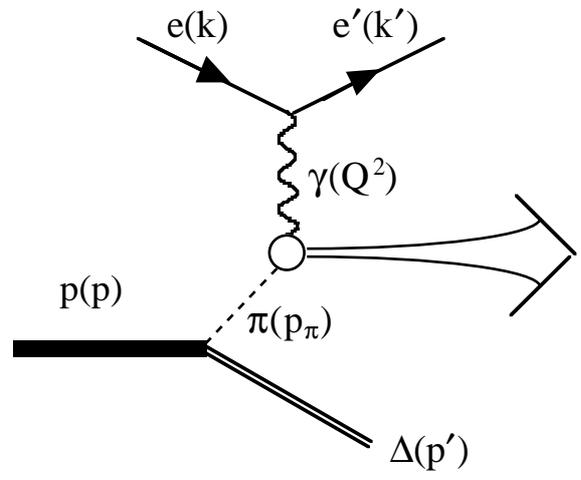
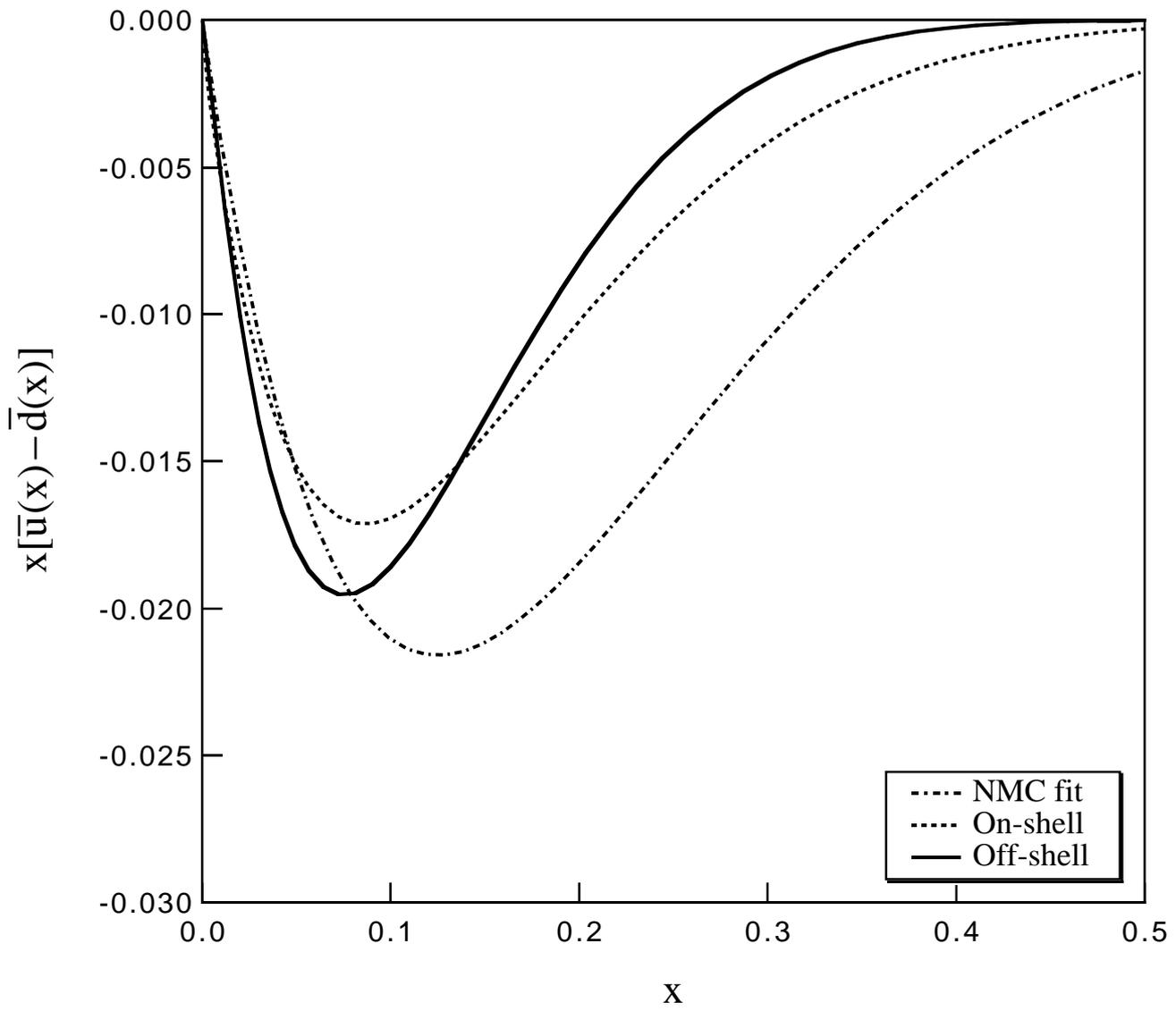
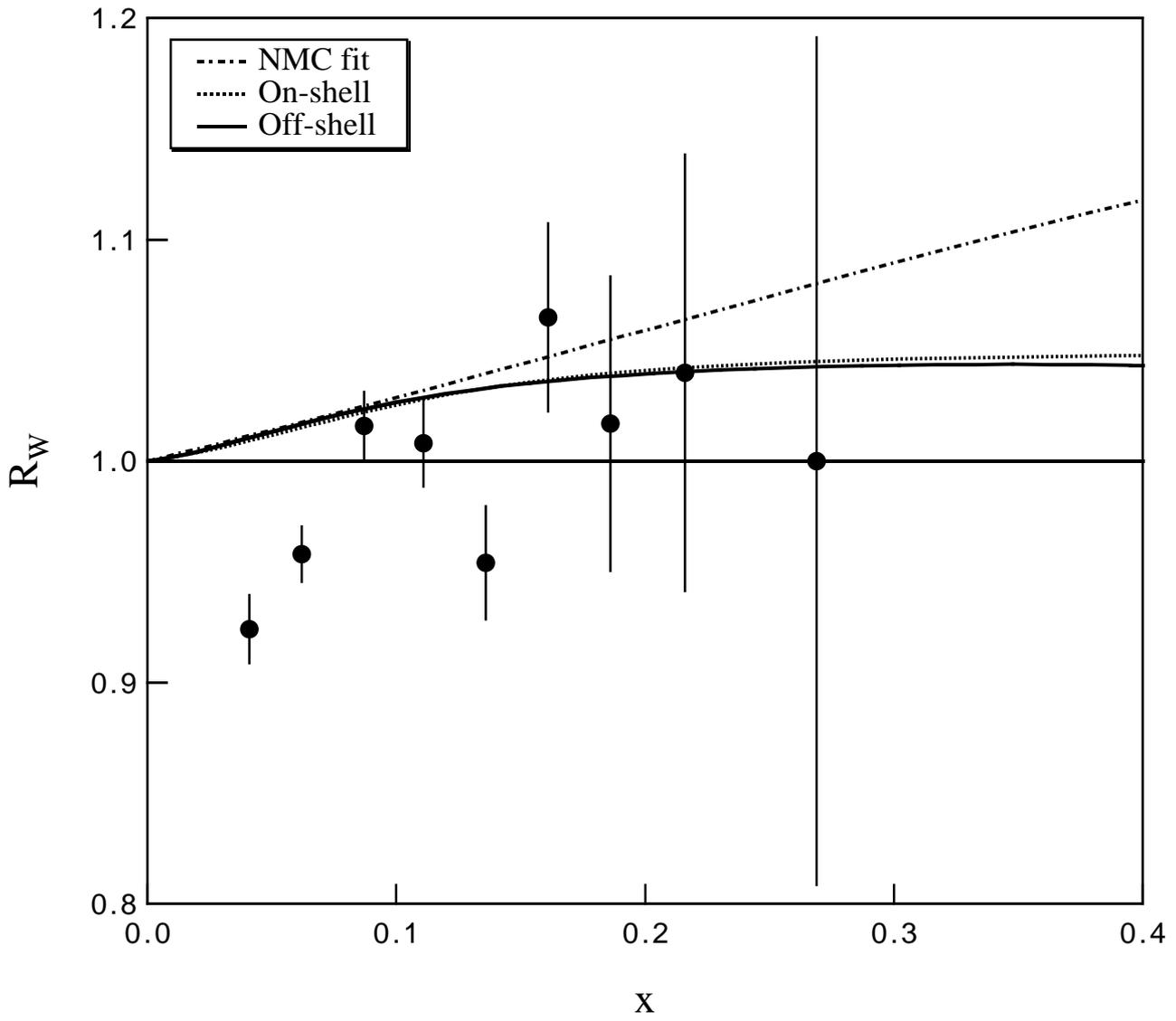
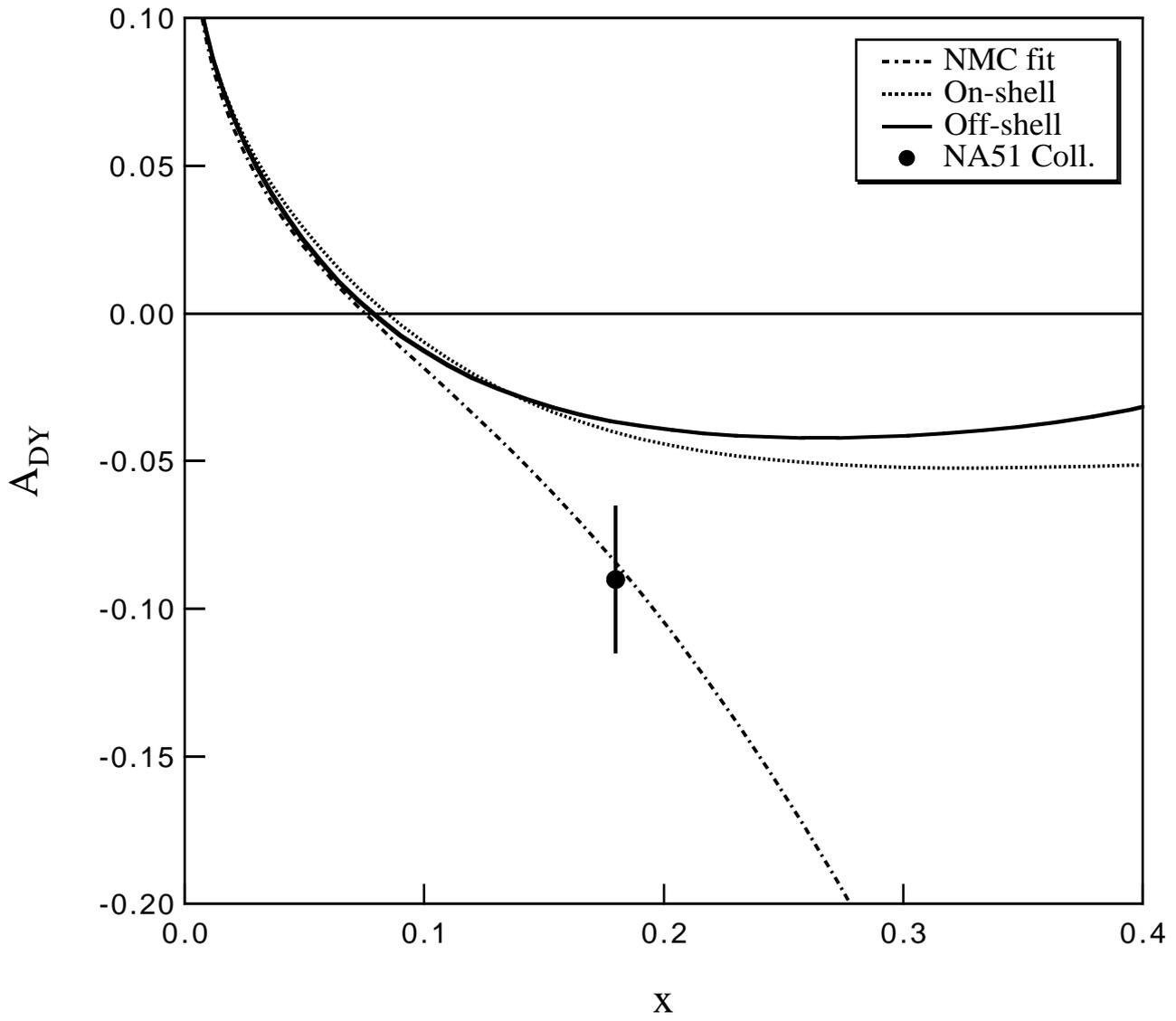
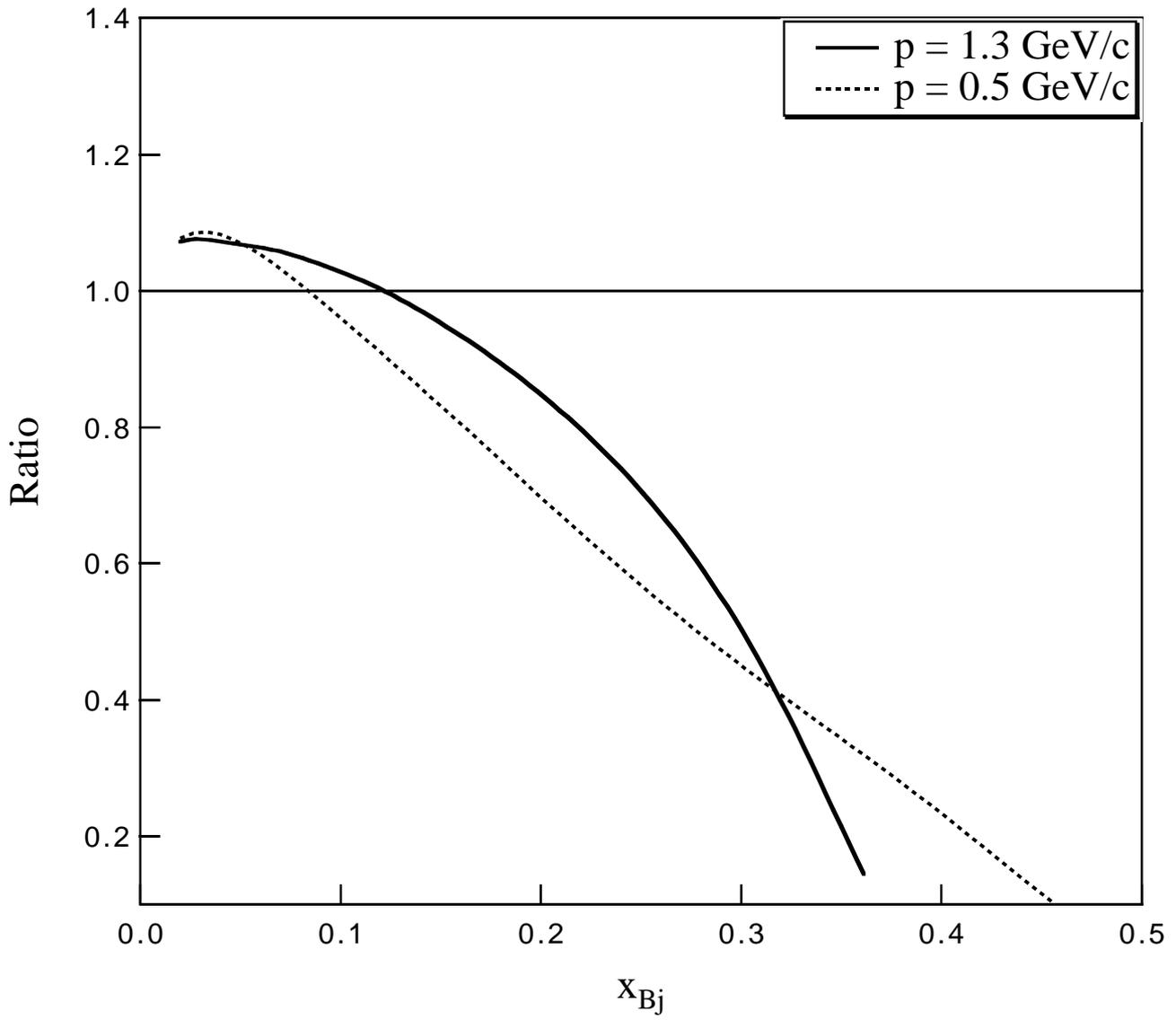
\end{document}